# The comparison of Wiktionary thesauri transformed into the machine-readable format[1]

Andrew Krizhanovsky

*Institution of the Russian Academy of Sciences St.Petersburg Institute for Informatics and Automation RAS*

Phone: +7 (812) 328-80-71

Fax: +7 (812) 328-44-50

andrew dot krizhanovsky@gmail.com

http://code.google.com/p/wikokit/

Wiktionary is a unique, peculiar, valuable and original resource for natural language processing (NLP). The paper describes an open-source Wiktionary parser: its architecture and requirements followed by a description of Wiktionary features to be taken into account, some open problems of Wiktionary and the parser. The current implementation of the parser extracts the definitions, semantic relations, and translations from English and Russian Wiktionaries. The paper's goal is to interest researchers (1) in using the constructed machine-readable dictionary for different NLP tasks, (2) in extending the software to parse 170 still unused Wiktionaries. The comparison of a number and types of semantic relations, a number of definitions, and a number of translations in the English Wiktionary and the Russian Wiktionary has been carried out. It was found that the number of semantic relations in the English Wiktionary is larger by 1.57 times than in Russian (157 and 100 thousands). But the Russian Wiktionary has more "rich" entries (with a big number of semantic relations), e.g. the number of entries with three or more semantic relations is larger by 1.63 times than in the English Wiktionary. Upon comparison, it was found out the methodological shortcomings of the Wiktionary.

## 1. INTRODUCTION

The creation of machine-readable dictionaries is an important step in the road of the automatic text processing. Machine-readable dictionaries and Wikipedias, and Wiktionaries are heavily used in different disciplines, including ontology building (Wandmacher et al. 2007), machine translation (Etzioni et al. 2007; Muller and Gurevych 2008), automatic text simplification (Napoles and Dredze 2010), image

[1] See Russian version of this paper: http://scipeople.ru/publication/99331/

search (Etzioni et al. 2007), and word sense disambiguation (Krovetz and Croft 1989).

Wiktionary is a unique resource and it could be useful for a wide variety of NLP tasks. But it cannot be used directly. There is a need to develop a software tool which makes it possible to convert the Wiktionary articles into a more suitable form for computer manipulation and processing, such as machine-readable dictionary (MRD).

This conversion software presented in the paper is a parser of the Wiktionary. The parser software is released under an open source license agreement (GPL), to facilitate its dissemination, modification and upgrades, to draw researchers and programmers into parsing other Wiktionaries, not only Russian and English.[2]

The specificity of Wiktionary is that it is created by the community of enthusiasts, and it is probably that not all of them are professional lexicographers. The structure of a dictionary entry is gradually, but constantly changing, since community experts regularly discuss and work out new and better rules. Also it should be taken into account that Wiktionary is permanently growing in number of entries and in the scope of languages. Now English Wiktionary contains about 740 different languages, and the parser recognizes 540 language codes.

Thus, the parsing of this linguistic resource makes high and specific demands on the software to be developed. These requirements are described in the next section.

The third section is concerned with the architecture of the Wiktionary parser. The comparisons of the main properties of Wiktionaries and their thesauri are presented in the fourth section. Discussion and related work conclude the paper.

## 2. REQUIREMENTS AND SOLUTIONS

Before the creation of the Wiktionary parser, the open-source software for the Wikipedia data extraction, the wikitext parsing, the indexing of Wikipedia texts (Krizhanovsky 2008), the search for related terms by analyzing Wikipedia internal links structure (Krizhanovsky 2006) was developed in our lab. This software can be used to create a parser, when the following conditions are met:

- Software must be written in the Java programming language.
- Wiktionary dump should be passed to the MySQL database.

---

[2] See http://code.google.com/p/wikokit

*Requirement*. Some requirements (problem statements) will be formulated for the successful design and implementation of a stable, functional, fast, modular and extensible parser.

> *Solution*. The required solution with implementation remarks will be described.

The requirements to the parser software code, to the structure of the parsed Wiktionary database and to the development process are listed below.

*Reliability and stability*. At this stage of the Wiktionary development there are no special MediaWiki features in order to control and prevent the malicious or erroneous input data. For example, the user can type a language code which does not exist or user can enter an extremely long definition sentence. Therefore the parser should properly treat the errors and defects.

> These requirements are satisfied due to the testing and the visualization (see below). *Unit testing*, as a method of extreme programming, is used, that is each non-trivial function is accompanied by one or more tests. The tests play an additional role by providing a sort of living documentation of the system.
>
> On June 2010, the parser source code contained 233 successfully passed tests and 29 failed tests. The dump of the Russian Wiktionary (2010, 300 thousands entries) has already been processed without failures. During the analyzing of the English Wiktionary (2010, 1.5 million entries) the parser was stumbled over about 10 "difficult" articles, which cause failures.

*Flexibility*. The Wiktionary formatting rules are constantly improving and changing. But if some rule is changed then the format and structure of all 1.5 millions of entries, even with the help of an army of bots, will not be changed at once and simultaneously in order to conform to the new rules. Only those entries which were edited after an adoption of a new rule will have an entirely new structure. A great number of entries corresponds to the preceding and outdated formatting rules, and this sad state of affairs can last for years until a volunteer editor will fix it.[3] Then the parser should be flexible enough to satisfy several formatting standards and yet to extract data from Wiktionary articles.

> The testing helps again. The inputs of the unit tests are the parts of entries with different formatting standards.

*Visualization*. The visual examination of the data from the parsed Wiktionary database will prompt for the missing fields which have not been extracted by the parser. That is, there is a need in a tool to quickly glance at all the sections of a Wiktionary entry saved in the MRD or to see that there are some problems. The visualization will help to avoid the formulation of the tedious and low-level SQL queries.

> The application *wiwordik* was being developed at the same time that the parser was being created. *wiwordik* is a visual interface (in the JavaFX programming language) to the machine-readable dictionary. It allows searching by the name of a Wiktionary entry or the word from the translation section. All the information extracted from the corresponding Wiktionary entry for the given word is presented to the user (Fig. 1).

*Wiktionary* ++ (breadth-first growing). It should not be difficult for the developer to add modules for the parsing of new Wiktionaries. This extensive growing of the parser (with a minimum of the additional and repeated work to be avoided) requires a clear and unambiguous division of the parser code into two parts: (1) the *kernel*, i.e. the part, which does not depend on the language, and (2) the *language-dependent part*, which have to be written anew for each added Wiktionary.

> The current parser implementation already works with two Wiktionaries: Russian and English. The language (Russian or English) is one of the input parameters of the parser (Fig. 2, see “Input”).

---

[3] The Wiktionary entries created in different years have some distinctions, so that the professional Wiktionary editor can assess (as if a geologist determines the relative ages of rock strata) the average age of the entry. The editor assesses it and puts the entry in order.

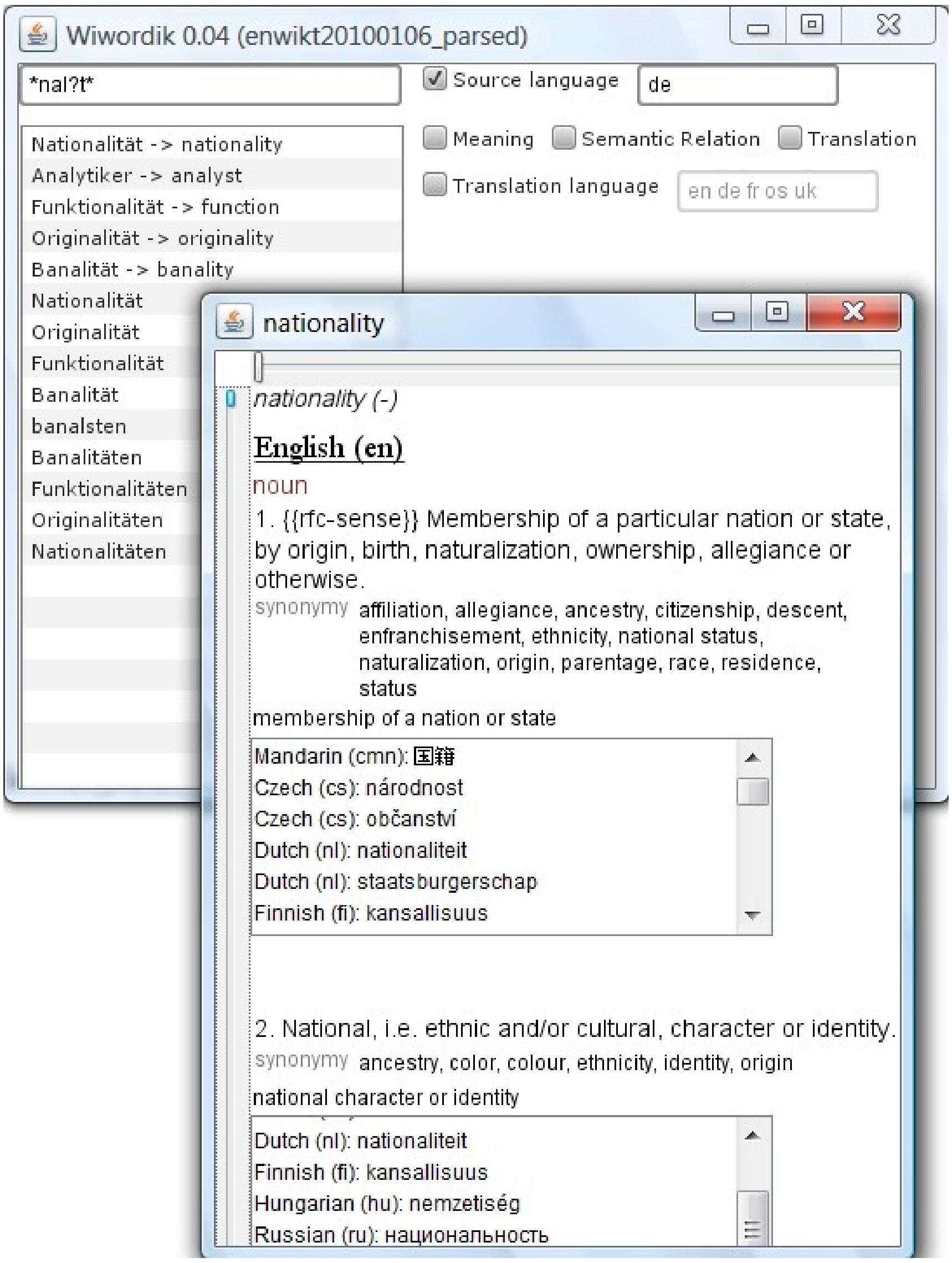

Fig. 1 The data about the entry “nationality” is extracted from the machine-readable dictionary

*Incremental approach* (depth-first growing). The incremental adding of new modules (in order to take in the new sections of a dictionary entry[4]) should not require significant code rewriting.

It is not reasonable to try to extract the data from all sections of an entry at once. Today only the definition, semantic relations and translations are extracted from the Wiktionary.

Apparently, the parser has been already developed on the foundation of an incremental approach. That is, first of all, the text of a Wiktionary entry is split into high-level parts (*Language*, in accordance with WT:ELE rules[5]), then the language sections are further split into *Etymology* subsections, then – *Parts of speech*, and so on. Thus, if there is a need, for example, to extract data from the *Pronunciation* section, then it is known which part of the parser code should be extended.

*Integration*. There is an important task to integrate the data extracted from different Wiktionaries into a common database. This integration is needed since dictionaries are built by hand and every Wiktionary contains unique data which are absent in other Wiktionaries.

It is an open question. On the one hand, our parsed Wiktionary databases (i.e. the database of the machine-readable dictionary) have the same structure (Fig. 3), so a practical question is how to merge the two databases into a single database. But, on the other hand, the different Wiktionaries may contain duplicate, contradictory, and inconsistent data, so the dictionary merging becomes an interesting theoretical question.

*Speed*. There is an important need in a rapid parsing of the Wiktionary dump in a reasonable time. Since the Wiktionary is constantly growing, the regular and frequent parsing of the dump is required in order to support the MRD in the actual state.

About 100-150 thousands of entries are parsed per day. One and a half million English Wiktionary entries are parsed during about 10 days (PC with 2.4 GHz Core 2 DUO and 3 GB of RAM).

---

[4] See the list and the structure of the Wiktionary entry subsections on the Wiktionary help page: http://en.wiktionary.org/wiki/Wiktionary:Entry_layout_explained

[5] See http://en.wiktionary.org/wiki/Wiktionary:ELE

The using of table indexes speeds up the parser significantly. The tables engaged in a search process should have indexes. The search is executed during the parsing, e.g. with the purpose to check whether there exists the word in the MRD database before adding it to the database.

# 3. ARCHITECTURE

Three rectangles drawn in the figure (Fig. 2) denote the following: the data processed by the program (parser), the program's input parameters and the modules of the program.

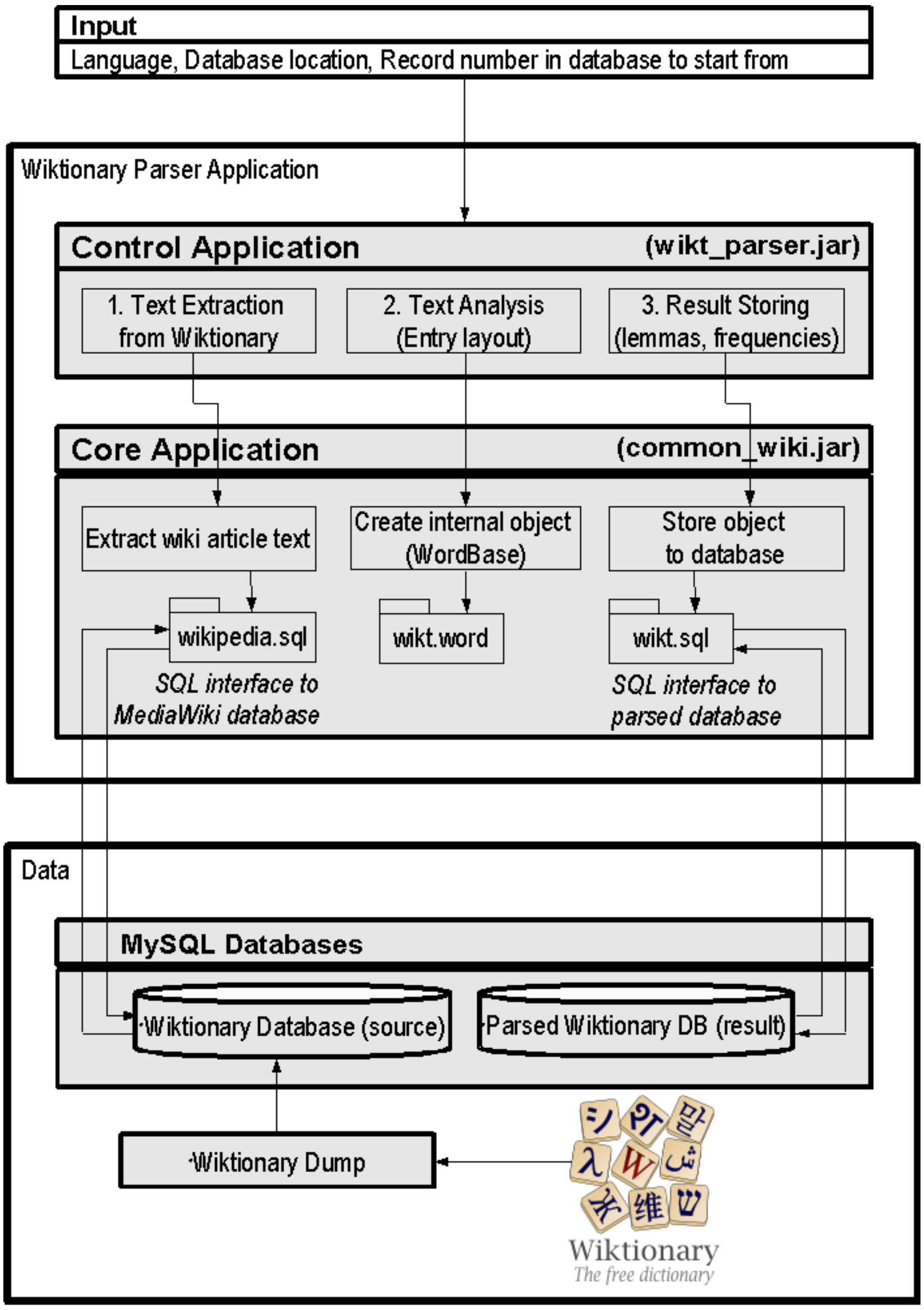

Fig. 2 Wiktionary parser architecture

*Data*. The main part of the program architecture is the Wiktionary. There are tens of voluntary editors of the Russian Wiktionary and several hundreds of unpaid editors of the English Wiktionary which work like mad. Due to this, the Wiktionaries grow gradually and develop by inner rules. Routinely, automatically, and one after the other the dumps of all Wikipedias and Wiktionaries are created and uploaded to the Wikimedia Foundation site.

Just one of these dumps uploaded to the local MySQL database is a source material, which will be processed by the parser in order to get the desired result, i.e. the parsed Wiktionary database (Fig. 2). In this resulting database the dictionary data are strongly structured, that is every lexicographic entity (definition, translation, etc.) has its own table in the database (Fig. 3).

*Parameters*. The meaning of the parameters is as follows (“*Input”*, the top part of the Fig. 2):

- *the native language* is the main language for the Wiktionary, e.g. Russian in the Russian Wiktionary;
- *the configuration parameters* to access to both databases (the source and the receiver databases);
- *the number of the record* in the source database, which give the possibility to stop and start again the parsing from this record.

The parser treats a Wiktionary article in three steps:

*(1) Extraction*. The title and the text of an article from the Wiktionary are extracted. This functionality has been implemented in the Java package *wikipedia.sql*.

*(2) Analysis*. The text of an article is analyzed. Many times (during the pass) various regular expressions are used to extract the desired information from the text. This extraction and analysis are possible only due to the known structure of the article and due to the applying of the templates.[6] The more strict and rigid structure of the entry is adopted by a community of wiki editors, the more simple and reliable will the parser’s algorithms be. The more number of templates widely used in the Wiktionary is, the more easy to extract the structured data from it.

The known project DBpedia (Bizer et al. 2009) also relies on the templates in the Wikipedia to extract the data from the encyclopedia.

---

[6] See http://en.wiktionary.org/wiki/Wiktionary:Templates

During the analysis of an article, a temporary intermediate Java object (the class *WordBase* in a Java package *wikt.word*) is created. The hierarchy of subclasses of this class corresponds to the structure of the Wiktionary article (the source) and to the tables in the created database (the receiver). The successful filling of this object with the extracted data is a prerequisite for the next step.

*(3) Saving*. The created object of the class *WordBase* (with all fields and all subclasses) is saved to the MRD database (see the Parsed Wiktionary database, Fig. 2).

The database layout of the MRD is presented in Fig. 3. This database is filled by the parsed data from Wiktionary.

In comparison with the previous publication (Krizhanovsky and Feiyu Lin 2009) there is one more table, which is denoted as *index_native* in the database scheme (Fig. 3). This table contains a list of native language words, e.g. English words in the English Wiktionary or Russian words in the Russian Wiktionary.

Index tables for other languages (which are not presented in the database layout) have the names *index_XX*, where XX is a language code. At this moment 540 index tables are created and filled by the parser, since the parser "knows" this number of language codes. But it is necessary to add still more language codes to the parser code (about several hundreds), in order to recognize all languages in the English Wiktionary.

These index tables have been added in order to speed up the search for words in the program *wiwordik*. Before all Wiktionary entries were stored in the huge table *page*. Now the table page is divided into several tables *index_XX*, one table for each language XX. This makes it possible to list (in a relatively fast manner) words of the given language in the program *wiwordik* (Fig. 1).

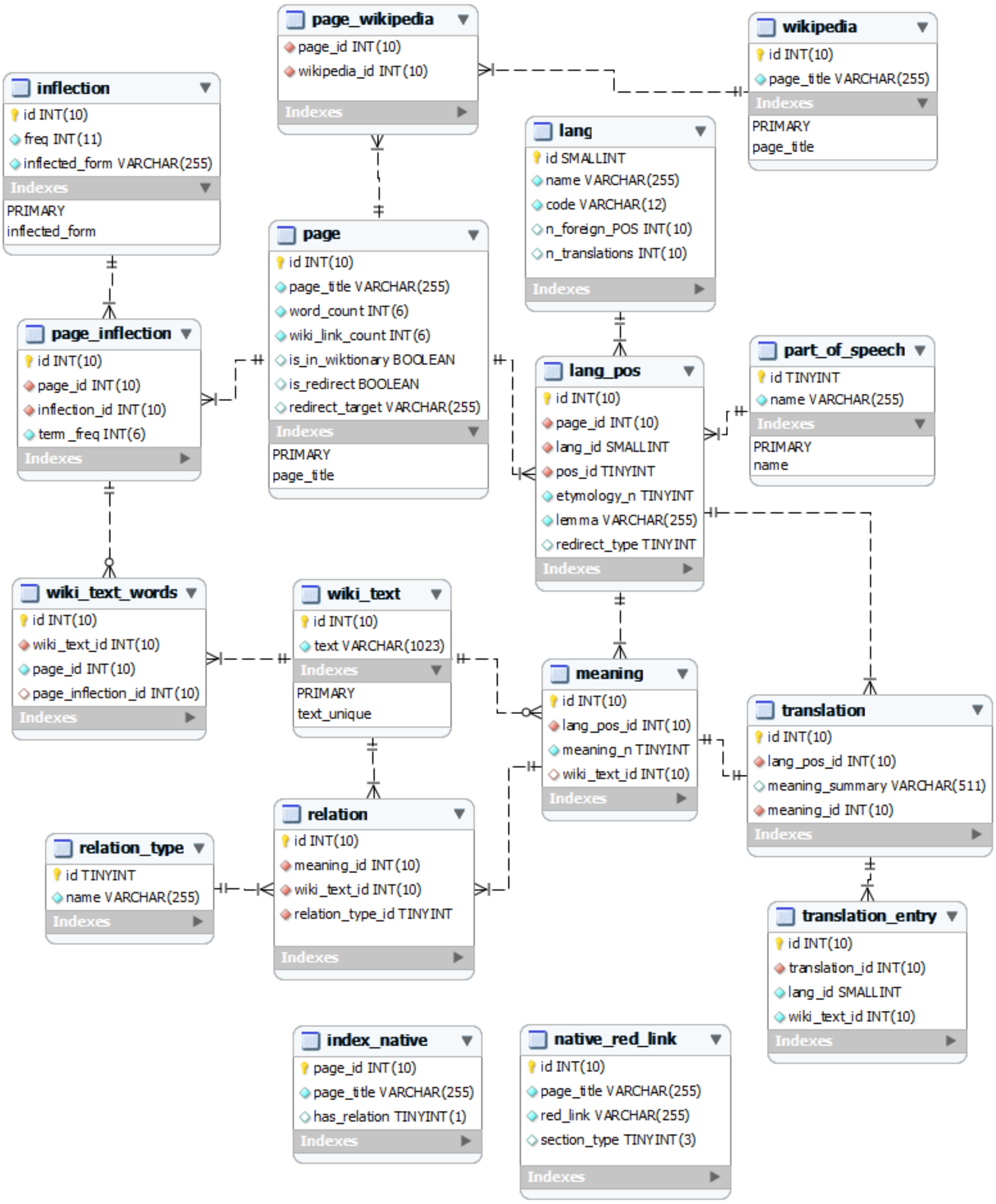

Fig. 3 Tables and relations in the database of the machine-readable dictionary (filled by the parsed data from Wiktionary)

# 4. EXPERIMENTS

During the experiments the databases of the machine-readable dictionaries were created. It allows comparing the different properties of Wiktionaries. The dump of the English Wiktionary (denoted as *enwikt*) as of Jan 6 2010 and the dump of the Russian Wiktionary (denoted as *ruwikt*) as of Apr 5 2010 were the source data for our experiments.

### 4.1 Comparison of the main properties of Wiktionaries

The main, from our point of view, properties of Wiktionaries and the sizes of the built machine-readable dictionaries are presented in the Table 1. The comparison of the Wiktionaries only (the section A in the table) shows that the number of pages and active users is about 7 times larger in the English Wiktionary than in the Russian Wiktionary. But there is nearly equal the average number of edits per page (about 5) in both dictionaries.

The percentage ratio of a number of entries with semantic relations to a number of content pages (Table 1, row 18) is almost twice as large for the Russian Wiktionary (10.7 %) compared to the English Wiktionary (5.8 %).

It was conducted a quantitative study of the *native language entries*, i.e. English language entries in the English Wiktionary were compared with Russian language entries in the Russian Wiktionary.

1. The number of semantic relations (Table 1, row 16) in the Russian Wiktionary between Russian words (84 thousands) is twice as large as that of the English Wiktionary (44 thousands).
2. The English language entries are the fifth part of all entries (18.3 %) in the English Wiktionary (row 17). The Russian language entries in the Russian Wiktionary constitute a much larger part; it is more than a half of all entries (53.7 %). So, in spite of the fact that one of the goals of *both* Wiktionaries is a "multilingual dictionary", the Russian Wiktionary is more monolingual at this moment.
3. The average number of semantic relations per entry (row 19) in the Russian Wiktionary (0.65) is larger than five times the number in the English Wiktionary (0.14).

The ratios of the sizes of the built machine-readable dictionary of the English Wiktionary to the Russian Wiktionary are in the range 3-8 (Table 1, rows 5-13). The more significant difference (in 12.7 times) exists in the size of the table "*meaning*" (See discussion of the possible reasons in the section *5.2 Number of meanings, word forms, lemmas, and sly bots*).

Table 1. The main parameters of the English Wiktionary, Russian Wiktionary and the created machine-readable dictionaries

| N | Property | English (en) | Russian (ru) | en / ru |
|---|---|---|---|---|
| | (A) Wiktionaries statistics (as of May 13, 2010) | | | |
| 1 | Content pages | 1 721 584 | 241 573 | 7.13 |
| 2 | Page edits since Wiktionary was set up | 9 230 581 | 2 529 788 | 3.65 |
| 3 | Average edits per page | 4,96 | 4,8 | 1.03 |
| 4 | Active users | 1082 | 151 | 7.17 |
| | (B) Machine-readable dictionaries filled by data from Wiktionaries, the size of tables (Fig. 3) | | | |
| | The date of the Wiktionary dump | Jan 6 2010 | April 5 2010 | – |
| | The table of MRD (and comment)[7] | | | |
| 5 | page | 1 721 798 | 456 138 | 3.77 |
| 6 | **relation** (number of semantic relations) | 157 198 | 100 121 | **1.57** |
| 7 | lang_pos | 1 732 162 | 374 257 | 4.63 |
| 8 | wiki_text | 2 151 393 | 275 530 | 7.81 |
| 9 | wiki_text_words | 3 356 231 | 310 398 | 10.81 |
| 10 | *meaning* | *2 158 845* | *170 313* | *12.68* |
| 11 | inflection | 205 219 | 23 208 | 8.84 |
| 12 | **translation** (total translation boxes, i.e. number of translated meanings of words) | 59 321 | 38 306 | **1.55** |
| 13 | translation_entry | 373 008 | 189 844 | 1.96 |
| 14 | Words (pairs: language & part of speech) with semantic relations | 100 268 | 25 747 | 3.89 |
| 15 | Words (pairs: language & part of speech) in native language | 315 343 | 129 669 | 2.43 |
| 16 | Number of semantic relations between words in native language | **43 814** | **83 968** | **0.52** |
| | ( C) Statistics calculated from (A) and (B) | | | |
| 17 | Words (pairs: language & part of speech) in native (main) language / Content pages [(15) / (1)], % | 18.32 | 53.68 | 0.34 |
| 18 | Words with semantic relations / Content pages [(14) / (1)], % | 5.82 | 10.66 | 0.55 |
| 19 | Average number of semantic relations for entries in native language [(16) / (15)] | 0.14 | 0.65 | 0.21 |

[7] The English Wiktionary statistics are available at http://en.wiktionary.org/wiki/User:AKA_MBG/Statistics:Parameters_of_the_database_created_by_the_Wiktionary_parser and Russian Wiktionary at http://ru.wiktionary.org/wiki/Участник:AKA MBG/Статистика:Размеры базы данных, созданной парсером Викисловаря

But the tables which have small differences are also interesting, e.g. the table "*relation*" (the ratio of sizes is 1.57) and the table "*translation*" (the ratio is 1.55). The following section will investigate the table "*relation*" in more detail.

In this paper the data extracted from the section "Translation" of the Wiktionary entry (and stored to the table "translation") are not analyzed, since there are a lot of language codes in the English Wiktionary to be added to the parser code (about several hundreds). Nevertheless, on the basis of these preliminary data, the numbers of translations extracted from the English Wiktionary[8] and from the Russian Wiktionary[9] are available online.

### 4.2 The comparison of Wiktionary semantic relations

The table "*relation*" of the machine-readable dictionary binds a type of a semantic relation (the table "*relation_type*"), a certain meaning of the word (the table "*meaning*") and the wikitext (the table "wiki_text") which consists of synonyms, antonyms, etc. (Fig. 2).

The size of the table "*relation*" is the number of semantic relations extracted from Wiktionary. 157 thousands of such relations were extracted from the English Wiktionary, 100 thousands from the Russian Wikipedia.

With the help of MRD the number of semantic relations per dictionary entry was counted. The result is presented in the (Fig. 4). E.g. the English noun "*toe*"[10] contains 7 semantic relations spread across 6 types of semantic relations (synonyms, antonyms, hyponyms, holonyms, meronyms, coordinate terms).

It should be noted that the number of semantic relations in the Fig. 4 is calculated separately for each homonym in spite of the fact that homonyms are included in one Wiktionary entry. E.g. the first English noun in the entry "*paw*"[11] has 12 words within 4 types of semantic relations; the second noun has 7 words in 5 types of semantic relations.

---

[8] See http://en.wiktionary.org/wiki/User:AKA_MBG/Statistics:Translations

[9] See http://ru.wiktionary.org/wiki/Участник:AKA_MBG/Статистика:Переводы

[10] See http://en.wiktionary.org/wiki/toe

[11] See http://en.wiktionary.org/wiki/paw

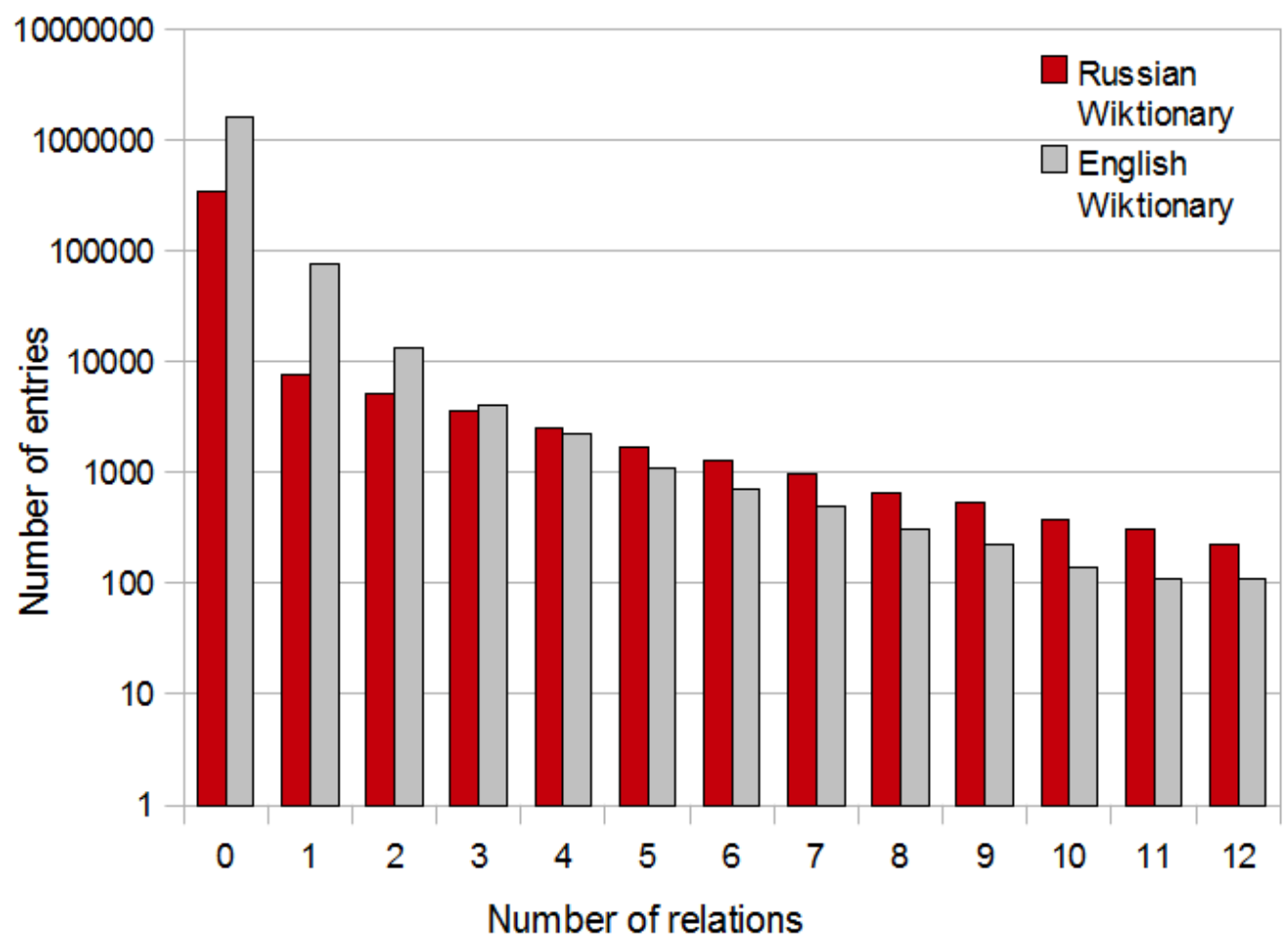

Fig. 4 Number of entries per number of relations (0-12) in the Russian Wiktionary and English Wiktionary

It can be concluded from the source data of the Fig. 4 that the number of entries with three or more semantic relations in the Russian Wiktionary is larger by 1.63 times than in the English Wiktionary. And vice versa, in the English Wiktionary the number of entries with one semantic relations is larger by 10 times, with two relations – by 2.5 times, and with three relations – by 1.12 times than in the Russian Wiktionary. So, the Russian Wiktionary has more "rich" entries, i.e. with a big number of semantic relations.

The Table 2 compares one more aspect of semantic relations: the number of types of semantic relations. E.g. the noun "*iron*"[12] contains 6 types (Synonyms, Hypernyms, Hyponyms, Meronyms, Holonyms, Coordinate terms) out of 9 possible.

The Table 2 shows that the numbers of words in the Russian Wiktionary which have 3, 4 types are significantly larger than in the English Wiktionary (about 60-170 times). Hypothesis explaining this fact see in the section 5.

The source data for this section (in a tabular form) and a list of words with many semantic relations are presented online.[13]

---

[12] See http://en.wiktionary.org/wiki/iron#Noun

[13] See http://ru.wiktionary.org/wiki/User:AKA_MBG/Статистика:Семантические_отношения and http://en.wiktionary.org/wiki/User:AKA_MBG/Statistics:Semantic_relations

Table 2. Number of words with different number of relation types in Russian Wiktionary (ru) and English Wiktionary (en)

| Types | Number of words | | ru / en | en / ru |
|---|---|---|---|---|
| | ru | en | | |
| 1 | 6254 | 16907 | 0.37 | **2.7** |
| 2 | 8167 | 3750 | 2.18 | 0.46 |
| **3** | 3215 | 53 | **60.66** | 0.02 |
| **4** | 844 | 5 | **168.8** | 0.01 |
| 5 | 45 | 1 | 45 | 0.02 |
| 6 | 6 | 1 | 6 | 0.17 |

It should be admitted that the comparison of semantic relations is not complete because of the Wikisaurus data of the English Wiktionary were not taken into account in this study. Our excuse for doing so is that (1) Wikisaurus is not presented in other editions of the Wiktionary, and (2) the Wikisaurus includes only English entries of the multilingual English Wiktionary. These are enough reasons to refrain from "teaching" the parser to extract data from Wikisaurus in the foreseeable future.

# 5. DISCUSSION

In this section the features of the Wiktionary as a linguistic resource and the Wiktionary characteristics essential for the parser are under consideration. Also the current state of the parser is discussed.

## 5.1 The problems and shortcomings of the parser

The difficulty of the parsing is that the goal of the Wiktionary is to create a *multilingual* free content dictionary. At this moment the parser recognizes only 479 pairs (language code – language name) out of 974 languages of the English Wiktionary[14].

The parser omits entries which are written in unknown languages. So the number of words, meanings and translations presented in the first table will increase after adding the missing language codes.

[14] See http://en.wiktionary.org/wiki/Wiktionary:Index_to_templates/languages#Template_table

### 5.2 Number of meanings, word forms, lemmas, and sly bots

In some wiktionaries the word forms (e.g. "*dog*" and "*dogs*") are described as different entries. The English Wiktionary entries describing these word forms (e.g. "*selects*", "*militias*") are created usually by bots, i.e. in an automatic manner. Of course, it is a good and convenient addition to the dictionary, but… While counting the number of all entries, such word form was taken as the full value entry. Perhaps, this is the reason of the big discrepancies between the ratios of parameters of two wiktionaries in Table 1, since in the Russian Wiktionary the rule is not to create a separate entry for each word form (all the more automatically), only for lemma.

Thus, one of the high priority tasks is to "teach" the parser to distinguish between the usual entry and the word form entry, which points to the entry with lemma (a kind of "soft redirect"). This distinction ability will allow comparing the real size of dictionaries.

### 5.3 Why the English Wiktionary is beaten by the Russian Wiktionary by a number of semantic relations?

The many parameters of the English Wiktionary outnumbered the parameters of the Russian Wiktionary by more than three to one ratio (3-8 times, see Table 1). However, the number of semantic relations in the English Wiktionary is larger only by a factor of 1.57. Moreover, the percentage ratio of a number of entries with semantic relations to a number of content pages is almost twice as large for the Russian Wiktionary compared to the English Wiktionary.

We can propose the following hypothesis in order to explain these findings: *the presence of empty headers* (subsections) in the section of semantic relations increases the completeness of this section of the entire Wiktionary.

The English Wiktionary rules do not recommend the presence of empty sections, e.g. see the entry "*dog*"[15] in the Fig. 5. But the policy of empty headers is supported by the community of the Russian Wiktionary, see the fragment of the entry «*собака*»[16] in the Fig. 6.

---

[15] See http://en.wiktionary.org/wiki/dog

[16] See http://ru.wiktionary.org/wiki/собака

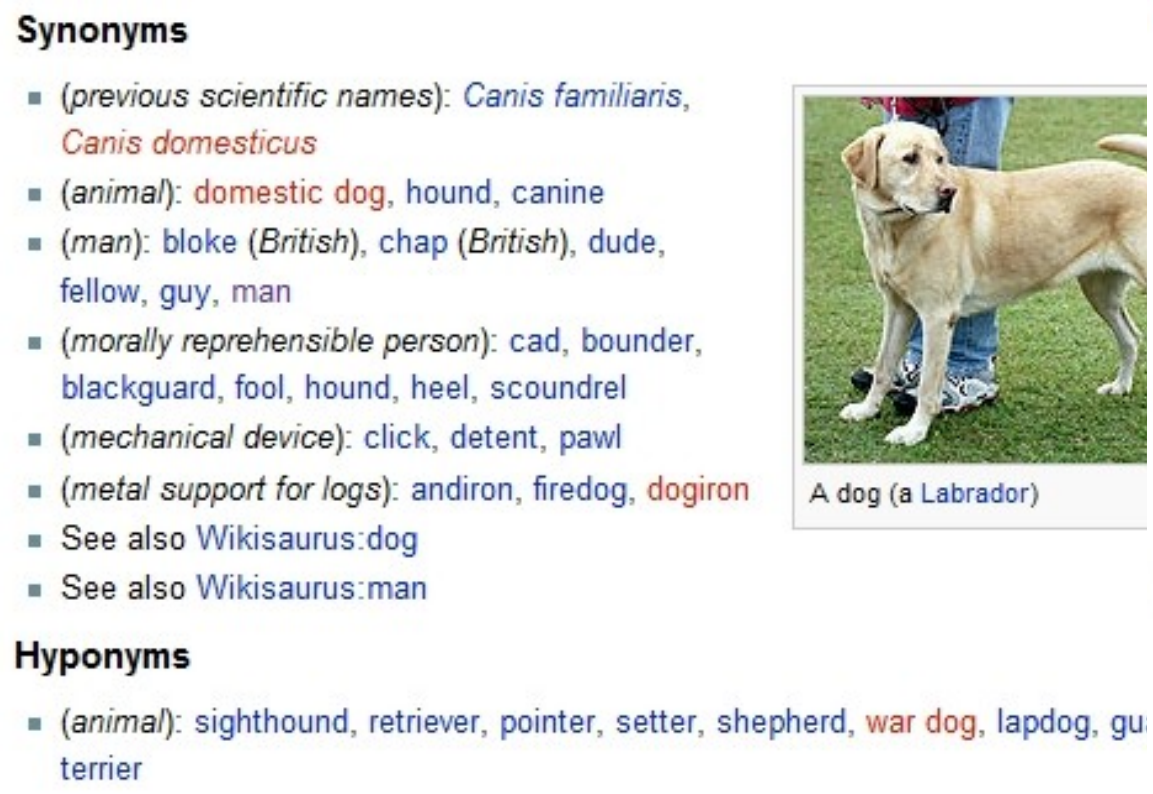

Fig. 5 En example of semantic relations sections in the English Wiktionary entry, there are no empty sections

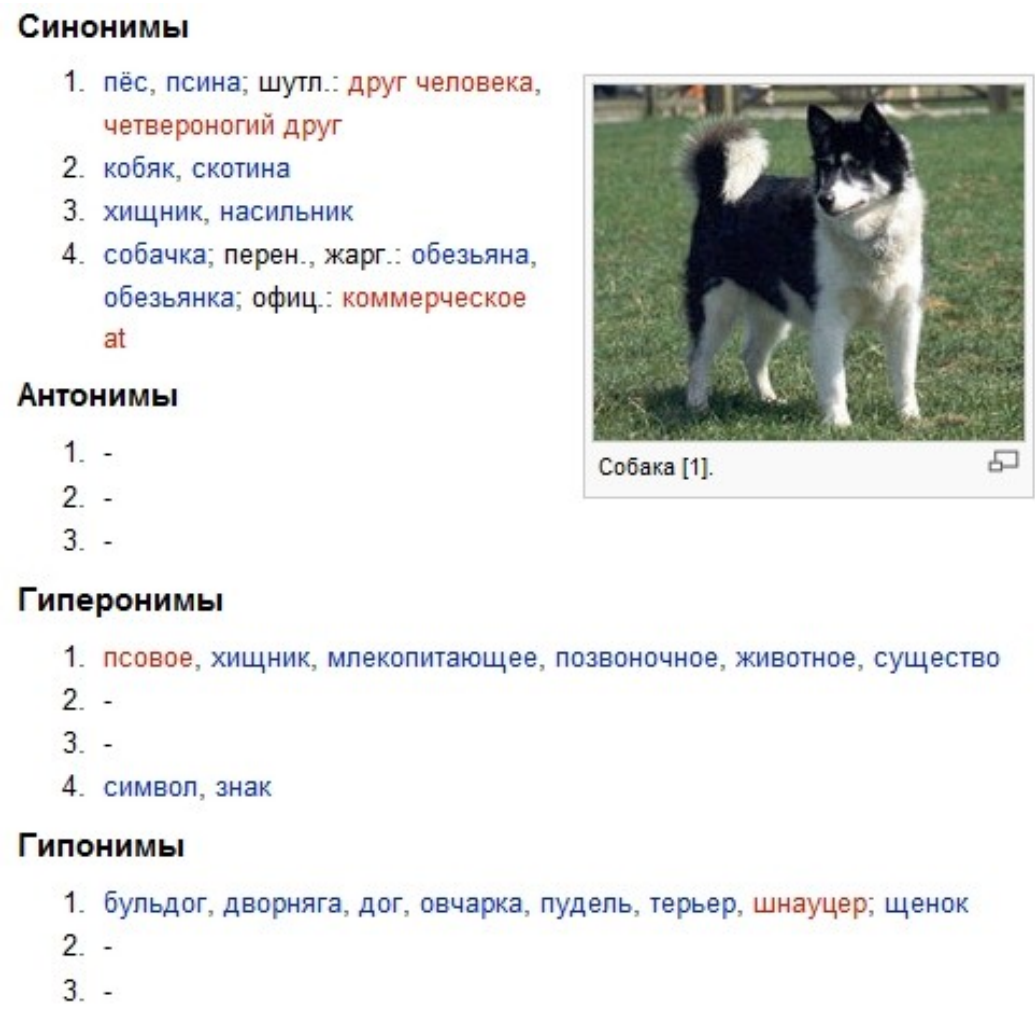

Fig. 6 En example of semantic relations sections in the Russian Wiktionary entry, one of which (Antonyms –Антонимы) is empty

It is our profound conviction that the appearance of empty sections provokes the editor to fill them and it simplifies the editing. It is evident that the presence of headers streamlines the process of editing and provides the ready-made entry layout. Thus, there is no need to take care of (1) the right order of headers, the order should conform to formatting rules, and (2) the correct number of equal signs “=”, which define a level of nesting in a wiki format.

Thus, the solution of the problem of relatively small number of the semantic relations (in comparison with other parameters of the English Wiktionary) consists in the *automatic creation of empty headers* of semantic relations subsections when the user creates new entry (so-called the New Entry Creation Wizard). That is, our recommendation is to adopt the successful practice of the Russian Wiktionary.

### 5.4 Explicit language names as a source of errors

From the parser developer’s point of view, the *manual* writing explicitly the language names (in order to define language of the entry, or in the translation section) is an unqualified and undoubted evil, since it gives room to make an error (misprint, misspelling). However, the editor will not be informed about this error.

Let us consider the fragments of the entries “bush”[17] and «ангел»[18] (Table 3). In the English Wiktionary the language of the word in question is written in the explicit form, in the Russian Wiktionary the special templates are used (rows 1-2, Table 3).

The definition of the language in the translation sections of the English Wiktionary is less naïve. It is used the special template {{t|}} (row 3). But there is an unnecessary and fallible duplication: the user has to write down explicitly the language name (e.g. Finnish) and it’s code (e.g. “fi”). In the Russian Wiktionary this problem was solved in a more elegant manner by using the huge template {{перев-блок|}}[19]. In this case the definition only of a language code is enough, e.g. “fi” for the Finnish or “ko” for the Korean (rows 3-4). Unfortunately for the parser, translations even without the template {{t|}} can be encountered in the English Wiktionary (row 4).

Table 3. The comparison of definitions of languages (explicitly or with the help of templates) in the English Wiktionary and the Russian Wiktionary

| N | English Wiktionary | Russian Wiktionary |
|---|---|---|
| Language of the entry | | |
| 1 | ==English== | = {{-en-}} = |
| 2 | ==Albanian== | = {{-sq-}} = |
| Translation section | | |
| 3 | * Finnish: {{t+\|fi\|pensas}} | \|fi=[[enkeli]] |
| 4 | * Korean: [[수풀]] (supul) | \|ko=[[천사]] |

Thus, from the parser development point of view, the Russian edition of the Wiktionary is more professional in this field, since (i) an entry language header and (ii) translation section use the rigid structure of templates. It requires more effort from the user (the editor needs to keep in mind the language codes), but it’s more fail-safe. So, if the user enters “et” instead of “es”, then immediately he will see the language name “Spanish” instead of “Estonian”. He will see and correct the misprint.

[17] See http://en.wiktionary.org/wiki/bush

[18] See http://ru.wiktionary.org/wiki/ангел

[19] See http://ru.wiktionary.org/wiki/Шаблон:перев-блок

## 5.5 Data of each Wiktionary is unique

It was mentioned more than once previously that each Wiktionary contains unique data, which is absent from other Wiktionaries.

This statement may be supported by the following fact. There are only 3.9 thousands of semantic relations between Russian words in the English Wiktionary, but there are 84 thousands in the Russian Wiktionary. And vice versa, there are 3.4 thousands of semantic relations between English words in the Russian Wiktionary, but there are 44 thousands in the English Wiktionary.

The following tasks will be left as exercises for the reader (or for the author) to support or refute this hypothesis by comparing the built machine-readable dictionaries:

- Find the degree of covering / intersection of words which belong to different languages, words with meanings, and the presence of semantic relations.
- Create a list of languages which are unique or almost unique, i.e. they are present only in one dictionary (Red List of Languages).
- Create two ordered lists: a list of languages which are better presented in one Wiktionary by different parameters (number of meanings, semantic relations), and a list for another Wiktionary.

## 6. RELATED WORK

Undoubtedly, the task of creating of machine-readable dictionaries existed long before Wiktionaries come on the scene (Krovetz and Croft 1989; Wilms 1990). However this amazing lexicographic resource, which is the Wiktionary, appeared only now.

*Manually* created thesauri, e.g. WordNet, have enjoyed considerable popularity in natural language processing. Thesauri which are filled *automatically* with data extracted from Wikipedia or from the Web are also actively used. It is most likely that the thesaurus presented in this paper (as a part of the built MRD) occupies intermediate position.

Not only the Wiktionary, but also the Wikipedia could be considered as a thesaurus. There are special algorithms to extract semantic relations from Wikipedia. E.g. the hyponyms and hypernyms are extracted from the Japan Wikipedia (Sumida and Torisawa 2008). More types of semantic relations were retrieved from the English Wikipedia in order to construct an ontology (Herbelot and Copestake 2006).

Nevertheless, there is a small number of studies related directly to the Wiktionary. A rare example is the paper (Zesch and Mueller 2008) which describes application programming interfaces for Wikipedia and Wiktionary (English and German Wiktionaries).

There is a number of works devoted to the comparison of Wiktionaries with other thesauri. In our previous research (Krizhanovsky and Feiyu Lin 2009) the related terms search based on the Russian Wiktionary was compared to WordNet based algorithms. The WordNet won.

The comparative study of the three resources German Wiktionary, GermaNet and OpenThesaurus that analyzes both topological and content related properties is presented in the paper (Meyer and Gurevych 2010). It was revealed that the German Wiktionary contains the lowest number of semantic relations (157 thousands, June 2009).

The Wiktionary in turn can be used to construct other thesauri. Thus, for example, French and Slovene WordNet were built by data extracted from different resources including French, Slovene, and English Wiktionary (Fiser and Sagot 2008).

## Conclusion

The architecture of the extensible and modular Wiktionary parser was developed. The modules for the extraction of three types of data from the Wiktionary entries (i.e. data of three subsections) were implemented: words meanings, semantic relations and translations. These modules were adapted to the English Wiktionary and the Russian Wiktionary, since formatting rules and the structure of the entry are different. The extensibility of the modular architecture enables (without significant rewriting of the previous code):

- to add some modules in order to extract data from other subsections of the entries, e.g. a Pronunciation, an Etymology, etc.;
- to adapt the existing modules in order to extract data from other language editions of Wiktionary, e.g. the French Wiktionary, the Chinese Wiktionary, etc.

The comparison of a number and types of semantic relations, a number of definitions, and a number of translations in the English Wiktionary and the Russian Wiktionary has been carried out. At that, some interesting findings have been discovered:

(1) The number of semantic relations in the English Wiktionary is larger by 1.57 times than in Russian (157 and 100 thousands).

(2) The percentage ratio of a number of entries with semantic relations to a number of content pages is almost twice as large for the Russian Wiktionary (10.7 %) compared to the English Wiktionary (5.8 %).

(3) It was conducted a quantitative study of the *native language entries*, i.e. English language entries in the English Wiktionary were compared with Russian language entries in the Russian Wiktionary.

   a. The number of semantic relations in the Russian Wiktionary between Russian words (84 thousands) is twice as large as that of the English Wiktionary (44 thousands).

   b. The English language entries are the fifth part of all entries (18.3 %) in the English Wiktionary. The Russian language entries in the Russian Wiktionary constitute a much larger part – it is more than a half of all entries (53.7 %). So, in spite of the fact that one of the goals of *both* Wiktionaries are "multilingual dictionary", the Russian Wiktionary is more monolingual at this moment.

   c. The average number of semantic relations per entry in the Russian Wiktionary (0.65) is larger than five times the number in the English Wiktionary (0.14).

There are many attractive ways to develop the parser and applications based on it. But, in the first place, the Graphical User Interface for the created machine-readable dictionary based on data from the English Wiktionary should be developed. This interface for the Russian Wiktionary is ready and available online.[20]

---

[20] See the program *wiwordik* based on the data extracted from the Russian Wiktionary: http://code.google.com/p/wikokit/wiki/wiwordik

## Acknowledgements

The paper is due to the research carried out as part of projects funded by grants 08-07-00264, 09-07-00436, 09-07-00066, and 09-07-12111 of the Russian Foundation for Basic Research, and project 213 of the research program "Intelligent information technologies, mathematical modeling, system analysis and automation" of the Russian Academy of Sciences.